\documentclass[12pt]{article}
\usepackage{amsfonts,amsmath,amssymb}
\usepackage{graphicx}
\usepackage{epic,eepic,color,graphicx}

\newfont{\twelvemsb}{msbm10 scaled\magstep1}
\newfont{\eightmsb}{msbm8}
\newfam\msbfam
\textfont\msbfam=\twelvemsb \scriptfont\msbfam=\eightmsb
\catcode`\@=11
\def\Bbb{\ifmmode\let\next\Bbb@\else
\def\next{\errmessage{Use \string\Bbb\space only in math mode}}\fi\next}
\def\Bbb@#1{{\fam\msbfam{{#1}}}}

\newcommand{\be}{\begin{equation}}
\newcommand{\ee}{\end{equation}}
\newcommand{\ba}{\begin{eqnarray}}
\newcommand{\ea}{\end{eqnarray}}

\begin{document}

\sloppy
\renewcommand{\thefootnote}{\fnsymbol{footnote}}
\newpage
\setcounter{page}{1} \vspace{0.7cm}
\begin{flushright}
17/04/08
\end{flushright}
\vspace*{1cm}
\begin{center}
{\bf Strong coupling for planar ${\cal N}=4$ SYM theory: an all-order result}\\
\vspace{1.8cm} {\large Davide Fioravanti $^a$, Paolo Grinza $^b$ and
Marco Rossi $^c$
\footnote{E-mail: fioravanti@bo.infn.it, pgrinza@usc.es, rossi@cs.infn.it}}\\
\vspace{.5cm} $^a$ {\em Sezione INFN di Bologna, Dipartimento di Fisica, Universit\`a di Bologna, \\
Via Irnerio 46, Bologna, Italy} \\
\vspace{.3cm} $^b${\em Departamento de Fisica de Particulas,
Instituto Galego de Fisica de Altas Enerxias (IGFAE), E-15782,
Universidade de Santiago de Compostela, Spain} \\
\vspace{.3cm} $^c${\em Dipartimento di Fisica, Universit\`a della
Calabria, I-87036 Arcavacata di Rende, Cosenza, Italy}
\end{center}
\renewcommand{\thefootnote}{\arabic{footnote}}
\setcounter{footnote}{0}
\begin{abstract}
{\noindent We propose a scheme for determining a generalised scaling
function, namely the Sudakov factor in a peculiar double scaling
limit for high spin and large twist operators belonging to the $sl(2)$
sector of planar ${\cal N}=4$ SYM. In particular, we perform
explicitly the all-order computation at strong 't Hooft coupling
regarding the first (contribution to the) generalised scaling
function. Moreover, we compare our asymptotic results with the
numerical solutions finding a very good agreement and evaluate
numerically the non-asymptotic contributions. Eventually, we illustrate
the agreement and prediction on the string side.}
\end{abstract}

\newpage

\section{On the set-up}
The $sl(2)$ sector of ${\cal N}=4$ SYM contains local composite
operators of the form
\begin{equation}
{\mbox {Tr}} ({\cal D}^s {\cal Z}^L)+.... \, , \label {sl2op}
\end{equation}
where ${\cal D}$ is the (symmetrised, traceless) covariant
derivative acting in all possible ways on the $L$ bosonic fields
${\cal Z}$. The spin of these operators is $s$ and $L$ is the
so-called 'twist'. Moreover, this sector would be described -- via
the AdS/CFT correspondence \cite {MWGKP} -- by string states on the
$\text{AdS}_5\times\text{S}^5$ spacetime with $\text{AdS}_5$ and
$\text{S}^5$ charges $s$ and $L$, respectively. Moreover, as far as
the one loop is concerned, the Bethe Ansatz problem is equivalent to
that of twist operators in QCD \cite{LIP, BDM}. \footnote{A deep
reason for that may be that one loop QCD still shows up conformal
invariance, albeit integrability does not seem to hold in full
generality (for instance, it apparently imposes aligned partonic
helicities).}

Proper superpositions of operators (\ref {sl2op}) have definite
anomalous dimension $\Delta$ depending on $L$, $s$ and the `t Hooft
coupling $\lambda = 8 \pi^2 g^2$:
\begin{equation}
\Delta = L+s+\gamma (g,s,L) \, , \label{Delta}
\end{equation}
where $\gamma (g,s,L)$ is the anomalous part. A great boost in the
evaluation of $\gamma (g,s,L)$ in another sector has come from the
discovery of integrability for the the purely bosonic $so(6)$
operators at one loop \cite {MZ}. Later on, this fact has been
extended to the whole theory and at all loops in the sense that, for
instance, any operator of the form (\ref {sl2op}) is associated to
one solution of some Bethe Ansatz-like equations and then any
anomalous dimension is expressed in terms of one solution \cite
{BES}. Nevertheless, along with this host of new results an
important limitation emerged as a by-product of the {\it on-shell}
(S-matrix) Bethe Ansatz: when the interaction range had been greater
than the chain length, then unpredicted {\it wrapping effects} would
have modified the same. More precisely, the anomalous dimension is
{\it in general} correct up to the $L-1$ loop in the convergent,
perturbative expansion, i.e. up to the order $g^{2L-2}$. Which in
particular implies, -- fortunately for us, -- that the {\it
asymptotic} Bethe Ansatz gives the right result whenever the $L
\rightarrow \infty$ limit is applied.

In this limit an important double scaling may be considered on both
sides of the correspondence \footnote{Actually, in string theory
(semi-classical calculations) the $\lambda\rightarrow +\infty$ limit
needs consideration before any other (cf. for instance \cite{FTT}
and references therein), thus implying a different limit order with
respect to our gauge theory approach (cf. below for more details).}:
\begin{equation}
s \rightarrow \infty \, , \quad L \rightarrow \infty \, , \quad
j=\frac {L}{\ln s}={\mbox {fixed}} \, . \label {jlimit}
\end{equation}
In fact, the relevance of this logarithmic scaling for the anomalous
dimension has been pointed out in \cite{BGK} within the one-loop SYM
theory and then in \cite{FTT} and \cite{AM} within the string theory
(strong coupling $g\gg 1$). Moreover, by describing the anomalous
dimension through a non-linear integral equation \cite{FRS} (like in other
integrable theories \cite{FMQR}), it has been recently conjectured
the Sudakov scaling \cite{AM}
\begin{equation}
\gamma (g,s,L)=f(g,j)\ln s + O((\ln s)^{-\infty}) \, ,
\end{equation}
thus indeed generalising to all loops the analogous result of \cite{BGK}.
\footnote{$O((\ln s)^{-\infty})$ means a remainder which goes faster
that any power of $\ln s$: $\lim\limits_{s \rightarrow \infty} (\ln
s)^k O((\ln s)^{-\infty}) =0, \forall \, k>0$.} Actually, this
statement was argued by computing iteratively the solution of some
integral equations and then, thereof, {\it the generalised scaling
function}, $f(g,j)$ at the first orders in $j$ and $g^2$: more
precisely the first orders in $g^2$ have been computed for the first {\it
generalised scaling functions} $f_n(g)$, forming
\begin{equation}
f(g,j)=\sum _{n=0}^{\infty} f_n(g)j^n \, .
\end{equation}
As a by-product, the reasonable conjecture has been put forward that
the two-variable function $f(g,j)$ should be analytic (in
$g$ for fixed $j$ and in $j$ for fixed $g$). In \cite{BFR2} similar
results have been derived for what concerns the contribution beyond
the leading scaling function $f(g)=f_0(g)$, but with a modification 
(see also \cite{BFR1} for the general idea) which has
allowed us to neglect the non-linearity for finite $L$ and to end-up with one
linear integral equation. The latter does not differ from the
BES one (which cover the case $j=0$, cf. the last of \cite{BES}), but for the
inhomogeneous term, which is an integral on the one loop root density.
In this respect, we have thought interesting the analysis of the 
next-to-leading-order (nlo)
term -- still coming, for finite $L$, from an asymptotic Bethe Ansatz --,
as the leading order $f(g)$ has been conjectured to be independent of $L$
or {\it universal} (penultimate reference of \cite{BES}, after the one loop
proof by \cite{BGK}).

For instance, the linearity in $L$ of this nlo term and other features
have furnished us the {\it inspiration} for the present calculations.
In this respect, we deem useful to briefly introduce in the next Section
a suitable modification of the aforementioned method such that it applies
to the regime  (\ref{jlimit}) (still for any $g$ and $j$). In principle, we
might in this way determine the generalised scaling function
$f(g,j)$, and also its {\it constituents} $f_n(g)$.  As a
consequence of the full range of $g$ we can work out not only
the weak coupling expansion, but also the exact strong coupling
evaluations (here only for the simplest case concerning $f_1(g)$), and to perform an
{\it all} $g$ numerical work. Eventually, we shall highlight the concise style of the present
publication, in that a next one with much more details shall be coming out
rather soon.

\section{Computing the generalised scaling function}
\setcounter{equation}{0}

Because of integrability in ${\cal N}=4$ SYM, we have been
finding useful \cite{NOI} to rewrite the Bethe equations as non-linear
integral equations, whose integration range is, even for massive
excited states (holes and complex roots), $(-\infty,+\infty)$ \cite{FMQR}. 
More recently \cite{BFR2, BFR1}, we have developed a new technique 
in order to cope with an integration
interval that may be kept finite, having in mind the peculiar case of
the $sl(2)$ sector with two holes external to all the roots \cite{BGK, BES, FRS}.
With reference in the entire Section to \cite{BFR2}, the non-linear integral
equation for the $sl(2)$ sector involves two functions $F(u)$ and
$G(u,v)$ satisfying linear equations. Splitting $F(u)$ into its one-loop
and higher loop contributions, $F_0(u)$ and $F^H(u)$ respectively, the latter
has been shown to satisfy (cf. (4.10) and (4.11) of \cite{BFR2}):
\begin{eqnarray}
\sigma _{H}(u)&=& -iL \frac {d}{du} \ln \left ( \frac {1+\frac {g^2}
{2{x^-(u)}^2}}{1+\frac {g^2}{2{x^+(u)}^2}} \right ) +
\nonumber \\
&-&  \frac {i}{\pi} \int _{-b_0}^{b_0} dv \frac {d}{du}\left [ \ln
\left ( \frac {1-\frac {g^2}{2x^+(u)x^-(v)} }{1-\frac
{g^2}{2x^-(u)x^+(v)}} \right )
+i \theta (u,v) \right ] [\sigma _0(v)+2\pi \delta_h (v)]+ \nonumber \\
&+&\int _{-\infty}^{+\infty} \frac {dv}{\pi} \frac {1}{1+(u-v)^2} [\sigma _{H}(v)+2\pi\eta_h(v)]- \label {sigmaeq} \\
&-& \frac {i}{\pi} \int _{-\infty}^{+\infty} dv \frac {d}{du} \left
[ \ln \left ( \frac {1-\frac {g^2}{2x^+(u)x^-(v)} }{1- \frac
{g^2}{2x^-(u)x^+(v)}} \right )+i \theta (u,v) \right ]  \sigma
_{H}(v) +O((\ln s)^{-\infty})\, , \nonumber
\end{eqnarray}
which holds when $s\rightarrow \infty$ once introduced in equation
(4.10) of \cite{BFR2} the Bethe roots densities \footnote{The actual 
positive densities would also have a factor $-1/(2\pi)$ in front \cite{BFR2}.} 
$\sigma _H(u) =\frac {d}{du}
F^H(u)$, $g$-dependent, and $\sigma _0(u)=\frac {d}{du} F_0(u)$, 
$g$-independent \footnote {The dependence $b_0(s)=s/2+O(s^0)$ comes out
from the normalisation of the density $\sigma _0(u)$ \cite {BFR2}.}.
Here the hole contributions given by 
$\delta_h (v)= (L-2)\delta (v)+ O(L^2/\ln s)$ \, and $\eta_h(v)=O(L^2/\ln s)$
depend on the logarithmic scaling to zero of the
internal holes. Moreover, the solution of (\ref{sigmaeq})
yields the anomalous dimension in the large spin limit while neglecting
the non-linear convolution term \footnote{We also have some numerical
evidence supporting this fact.}:
\begin{eqnarray}
\gamma (g,s,L)&=&-g^2\int _{-b_0}^{b_0}\frac {dv}{2\pi} \left [\frac
{i}{x^+(v)}-\frac {i}{x^-(v)}\right ] [\sigma _0(v)+2\pi \delta_h (v)] -\nonumber \\
&-& g^2\int _{-\infty}^{+\infty}\frac {dv}{2\pi} \left [\frac
{i}{x^+(v)}-\frac {i}{x^-(v)}\right ] \sigma_H(v) + O((\ln s)^{-\infty}) =
 \label{egs} \\
&=& \frac {1}{\pi} \hat \sigma _H(k=0) + O((\ln
s)^{-\infty}) \label {egs2}\, ,
\end{eqnarray}
where the last equality is a generalisation of the analogue in \cite{KL}.

As we are just neglecting terms smaller than any inverse logarithm,
$O((\ln s)^{-\infty})$, we can efficiently consider the double scaling limit (\ref {jlimit}).
Looking at (\ref{sigmaeq}), the r.h.s. is made up of seven terms, among which 
the first is explicit and the last uninfluential. On the contrary, the third
and the fifth integral reflect two different kinds of important hole 
contributions. Nevertheless, thanks to the quoted scaling
$\delta_h (v)= (L-2)\delta (v)+ O(L^2/\ln s)$ \, and $\eta_h(v)=O(L^2/\ln s)$, 
these integrals enjoy an expansion about $j=0$ (similar to the following one) 
contributing only with the 
Dirac delta function to $f_1(g)$, and the remainder (in particular the all 
$\eta_h(v)$) to the higher $f_n(g), \, n=2,3,\dots$\,. And last but not least, the second 
term determines the inhomogeneous (or forcing) term as a series about $j=0$ via the 
density $\sigma _0(u)$, by virtue of the {\it logarithmic expansion} \cite{BGK, FRS}
of the one-loop theory \cite{BFR2}:
\begin{equation}
-\frac {i}{\pi} \int _{-b_0}^{b_0} dv \frac {d}{du}\left [ \ln \left
( \frac {1-\frac {g^2}{2x^+(u)x^-(v)} }{1-\frac
{g^2}{2x^-(u)x^+(v)}} \right ) +i \theta (u,v) \right ] \sigma
_0(v)= \left[ \sum _{n=0}^{\infty}\phi _n(u)j^n \right] \ln s + O((\ln
s)^{-\infty}) \, .
\end{equation}
In fact, this series shall follow from the solution (in the scaling
(\ref{jlimit})) of the linear integral equation for the one
loop density $\sigma _0(u)$ (equation (4.9) of \cite {BFR2}), which determines
all the functions $\phi _n(u)$. As a consequence, the solution of (\ref{sigmaeq})
inherits the same form of the forcing term
\begin{equation}
\sigma _H(u)= \left[  \sum _{n=0}^{\infty} \sigma
_H^{(n)}(u)j^n \right]  \ln s  + O((\ln s)^{-\infty}) \, .
\end{equation}
Eventually, all the generalised scaling functions $f(g,j)$ and
$f_n(g)$ may be computed, in principle, via the equality (\ref{egs2}).
From now on, we will restrict ourselves to the first order beyond the $j=0$
theory \cite{BES}, i.e. $\sigma _H^{(1)}(u)$ and
$f_1(g)$.

\section{The first generalised scaling function}
\setcounter{equation}{0}

In \cite{BFR2} we have found a way to compute the Stieltjes 
integrals with measure induced by the one loop-density $\sigma _0(v)$ 
on the support $(-b_0,b_0)$ via an {\it effective} density  
$\sigma _0^{(b_0)}(v)$ on support $(-\infty,+\infty)$
\begin{equation}
\int _{-b_0}^{b_0} dv  \sigma _0(v) q(v)= \int _{-\infty}^{\infty} dv
\sigma _0^{(b_0)}(v) q(v) +O(1/\ln s ) \, , \label{1loop}
\end{equation}
where the Fourier transform of the function $\sigma _0^{(b_0)}(v)$ reads
explicitly
\begin{equation}
\hat \sigma _0^{(b_0)}(k)=-4\pi \frac {\frac {L}{2}-e^{-\frac {|k|}{2}}
\cos \frac {ks}{\sqrt {2}}} {2\sinh \frac {|k|}{2}}+2\pi (L-2) \frac
{e^{-\frac {|k|}{2}}}{2\sinh \frac {|k|}{2}}-4\pi \delta (k) \ln 2
\, ,
\end{equation}
and $q(v)$ is a suitable test-function. Therefore, we can compute exactly 
the functions $\phi_0(u)$ and $\phi_1(u)$ 
and thus write down the linear integral equations for $\sigma _{H}^{(0)}(u)$ and 
$\sigma _{H}^{(1)}(u)$ respectively. Of course, the equation for
$\sigma _{H}^{(0)}(u)$ is the BES equation \cite{BES}, whilst that about 
$\sigma _{H}^{(1)}(u)$ is a novelty. We may find still convenient to manipulate the 
Fourier transforms defining the even function
\begin{equation}
s(k)= \frac {2\sinh \frac {|k|}{2}}{2\pi  |k|}\hat\sigma _H^{(1)}(k)
\label {Sdef} \, ,
\end{equation}
which has the important property (induced by (\ref{egs2}))
\begin{equation}
s(0)=\frac {1}{2}f_1(g) \, . \label {sf_1}
\end{equation}
If we introduce for convenience the functions
\begin{equation}
a_r(g)=\int _{-\infty}^{+\infty}\frac {dh}{h} J_{r}({\sqrt
{2}}gh)\frac {1}{1+e^{\frac {|h|}{2}}} \, , \quad \bar a_r(g)=\int
_{-\infty}^{+\infty}\frac {dh}{|h|} J_{r}({\sqrt {2}}gh)\frac
{1}{1+e^{\frac {|h|}{2}}} \, ,
\end{equation}
and re-cast the integration inside the domain $k\geq 0$, the integral equation 
for $s(k)$ takes the form
\begin{eqnarray}
s(k)&=&\frac {1-J_0({\sqrt {2}}gk)}{k}-\sum _{p=1}^{\infty}2p \bar a_{2p}(g)\frac {J_{2p}({\sqrt {2}}gk)}{k}-\sum _{p=1}^{\infty}(2p-1) a_{2p-1}(g)\frac {J_{2p-1}({\sqrt {2}}gk)}{k}\nonumber \\
&+&2 \sum _{p=1}^{\infty}\sum _{\nu =0}^{\infty} \Bigl [ (-1)^{1+\nu}c_{2p+1,2p+2\nu+2}(g)a_{2p+2\nu+1}(g) \frac {J_{2p}({\sqrt {2}}gk)}{k}+\nonumber \\
&+& (-1)^{1+\nu}c_{2p,2p+2\nu+1}(g)a_{2p-1}(g) \frac {J_{2p+2\nu}({\sqrt {2}}gk)}{k} \Bigr ] - \nonumber \\
&-& 2  \sum _{p=1}^{\infty}2p  \frac {J_{2p}({\sqrt {2}}gk)}{k}\int _{0}^{\infty}dh \frac  {J_{2p}({\sqrt {2}}gh)}{e^h-1}s(h) - \label {seq2} \\
&-& 2  \sum _{p=1}^{\infty}(2p-1)  \frac {J_{2p-1}({\sqrt {2}}gk)}{k}\int _{0}^{\infty}dh \frac  {J_{2p-1}({\sqrt {2}}gh)}{e^h-1}s(h)  + \nonumber \\
&+&4   \sum _{p=1}^{\infty}\sum _{\nu =0}^{\infty} \Bigl [ (-1)^{1+\nu}c_{2p+1,2p+2\nu+2}(g) \frac {J_{2p}({\sqrt {2}}gk)}{k}\int _{0}^{\infty}dh \frac  {J_{2p+2\nu+1}({\sqrt {2}}gh)}{e^h-1}s(h) \nonumber \\
&+& (-1)^{1+\nu}c_{2p,2p+2\nu+1}(g) \frac {J_{2p+2\nu}({\sqrt
{2}}gk)}{k}\int _{0}^{\infty}dh \frac  {J_{2p-1}({\sqrt
{2}}gh)}{e^h-1}s(h) \Bigr ] \, . \nonumber
\end{eqnarray}
As anticipated, this equation shares the same kernel with BES equation \cite{BES}, but different forcing term. Its solution may be expanded in a series involving Bessel functions:
\begin{eqnarray}
s(k)&=&\frac {1-J_0({\sqrt {2}}gk)}{k}+\sum _{p=1}^{\infty}s_{2p}(g)\frac {J_{2p}({\sqrt {2}}gk)}{k}+\sum _{p=1}^{\infty}s_{2p-1}(g)\frac {J_{2p-1}({\sqrt {2}}gk)}{k} = \nonumber \\
&=& \sum _{p=1}^{\infty}\left (2+s_{2p}(g)\right )\frac
{J_{2p}({\sqrt {2}}gk)}{k}+\sum _{p=1}^{\infty}s_{2p-1}(g)\frac
{J_{2p-1}({\sqrt {2}}gk)}{k} \, . \label {smexp}
\end{eqnarray}
Now we want to derive linear equations for the coefficients $s_n$. Upon expressing 
the scattering factor coefficients via 
\begin{equation}
Z_{n,m}(g)=\int _{0}^{\infty} \frac {dh}{h} \frac {J_n({\sqrt
{2}}gh)J_m({\sqrt {2}}gh)}{e^h-1} \, ,
\end{equation}
in the form
\begin{equation}
c_{r,s}(g)=2 \cos \left [ \frac {\pi}{2}(s-r-1)\right ] (r-1)(s-1)
Z_{r-1,s-1}(g) \, ,
\end{equation}
equation (\ref {seq2}) decomposes into the infinite dimensional linear system
\begin{eqnarray}
S_{2p}(g)&=&2+2p \left ( -\bar a_{2p}(g)-2\sum _{m=1}^{\infty}Z_{2p,2m}(g)S_{2m}(g)+2\sum _{m=1}^{\infty}Z_{2p,2m-1}(g)s_{2m-1}(g) \right ) \nonumber \\
\label {Seq2} \\
\frac {s_{2p-1}(g)}{2p-1}&=&-a_{2p-1}(g)-2\sum
_{m=1}^{\infty}Z_{2p-1,2m}(g)S_{2m}(g)-2\sum
_{m=1}^{\infty}Z_{2p-1,2m-1}(g)s_{2m-1}(g) \nonumber \, ,
\end{eqnarray}
where we have re-defined for brevity
\begin{equation}
S_{2m}(g)=2+s_{2m}(g) \, .
\end{equation}

\section{Exact asymptotic expansion}
\setcounter{equation}{0}

At this point, we start looking for the solution of equations (\ref{Seq2}) 
for $g\rightarrow +\infty$ in the form of an asymptotic series, i.e.
\begin{equation}
s_{2m}(g)\doteq \sum _{n=0}^{\infty
} \frac {s_{2m}^{(n)}}{g^n} \, , \quad s_{2m-1}(g)
\doteq \sum _{n=0}^{\infty } \frac
{s_{2m-1}^{(n)}}{g^n} \, . \label {sexpan}
\end{equation}
Albeit we well know that the unique knowledge of the asymptotic expansion will 
never identify the function itself (contrary to the convergent weak-coupling 
series \cite{FRS, BFR2}), we will see in the following many favourable issues, 
besides the contact with semi-classical string theory results.
From the explicit solution of (\ref{Seq2}) up to the order $1/g^2$, we have 
guessed the form at all orders:
\begin{eqnarray}
&& s_{2m}^{(2n)}=2m \frac {\Gamma (m+n)}{\Gamma
(m-n+1)}(-1)^{1+n}b_{2n} \, , \quad
s_{2m-1}^{(2n)}=0 \, ; \  n \geq 0 \, , \quad m \geq 1 \, , \nonumber \\
&& \label {guess} \\
&& s_{2m}^{(2n-1)}=0 \, , \quad s_{2m-1}^{(2n-1)}=(2m-1)\frac
{\Gamma (m+n-1)}{\Gamma (m-n+1)}(-1)^n b_{2n-1} \, ;\ n \geq 1 \, ,
\quad m \geq 1 \, , \nonumber
\end{eqnarray}
with, implicitly, these expressions different from zero only if $n\leq m$. Moreover, the
coefficients $b_n$ satisfy a recursive relation whose solution is more easily written 
in terms of their generating function
\begin{equation}
b(t)=\sum _{n=0}^{\infty} b_n t^n \, ,
\end{equation}
which is then worth
\begin{equation}
b(t)=\frac {1}{\cos \frac {t}{\sqrt {2}}-\sin \frac {t}{\sqrt {2}}}
\, .
\end{equation}
Consequently, we may also derive the explicit values 
\begin{eqnarray}
b_{2n}&=&2^{-n}(-1)^{n} \sum _{k=0}^{n} \frac
{E_{2k}2^{2k}}{(2k)!(2n-2k)!} \, , \nonumber \\
&& \label {aeul} \\
b_{2n-1}&=&2^{-n+\frac {1}{2}}(-1)^{n-1} \sum _{k=0}^{n-1} \frac
{E_{2k}2^{2k}}{(2k)!(2n-2k-1)!} \, , \nonumber
\end{eqnarray}
where $E_{2k}$ are Euler's numbers. 
To summarise equations (\ref {sexpan}, \ref {guess}, \ref {aeul}), we
obtain the following strong coupling solution to the linear system (\ref{Seq2}):
\begin{eqnarray}
s_{2m-1}(g)&\doteq&- (2m-1)\sum
_{n=1}^m \frac {\Gamma (m+n-1)}{\Gamma (m-n+1)} \frac
{2^{\frac {1}{2}-n}}{g^{2n-1}}\sum
_{k=0}^{n-1} \frac {E_{2k}2^{2k}}{(2k)!(2n-2k-1)!} \, , \nonumber \\
&& \label {sm} \\
s_{2m}(g)&\doteq&- 2m\sum
_{n=0}^m \frac {\Gamma (m+n)}{\Gamma (m-n+1)} \frac
{2^{-n}}{g^{2n}}\sum _{k=0}^{n} \frac {E_{2k}2^{2k}}{(2k)!(2n-2k)!}
\, . \nonumber
\end{eqnarray}
As these series truncate, they are {\it asymptotic} in a very peculiar sense, 
namely they are correct up to non-expandable corrections which still 
go to zero (cf. below). From definition (\ref{smexp}) and property (\ref{sf_1}) we can read off
\begin{equation}
s(0)=\frac {g}{\sqrt {2}}s_1(g)=\frac
{1}{2}f_1(g) \, , \label {s1f1}
\end{equation}
which entails the exact asymptotic expression $f_1(g)\doteq -1$.

It is very interesting to provide a possible summation for
the expansions (\ref{sm}) (though finite!) by using an integral representation 
for the ratio of the two gamma functions and then by exchanging the sum over
$n$ with the integral, and finally summing up: this procedure provides these 
{\it new} functions of $g$ 
\begin{eqnarray}
\tilde s_{2m-1}(g)&=&-2(2m-1) \int _{0}^{\infty} \frac {dt}{t}\frac
{\sinh t}{\cosh 2 t} J_{2m-1}(2{\sqrt {2}}gt) \nonumber \\
&& \label {smint} \\
\tilde s_{2m}(g)&=&-4m \int _{0}^{\infty} \frac {dt}{t}\frac {\cosh
t}{\cosh 2 t } J_{2m}(2{\sqrt {2}}gt) \, . \nonumber
\end{eqnarray}
Instead of the first generalised scaling function we obtain
\begin{equation}
\tilde f_1(g)=-2 {\sqrt {2}} g \int _{0}^{\infty} \frac {dt}{t}\frac
{\sinh t}{\cosh 2 t} J_{1}(2{\sqrt {2}}gt)  \label {f1int} \, .
\end{equation}
From this integral representation we see that the asymptotic behaviour becomes
corrected by exponentially small terms, behaving as
$g^{1/2}{\mbox {exp}} (-\pi g/{\sqrt {2}})$. We will see in next
section that this is not quite the right behaviour because of the wrong power 
$g^{1/2}$. This is obviously connected to the wrong (convergent) weak
coupling expansion of (\ref{f1int}).

\section{Numerical solution: the non-asymptotic terms}

Remarkably, the linear system (\ref{Seq2}) shows up the same
matrix as in \cite{BBKS} for the BES equation, but a more involved forcing term. 
Explicitly, it may assume the form
\begin{eqnarray}
\label{matsys} s_p(g) = b_p(g) - \sum_{m=1}^{\infty}
(K^{(m)}_{pm}(g)+2 K^{(c)}_{pm}(g)) s_m(g)\, ,
\end{eqnarray}
where
\begin{eqnarray}
&& K^{(m)}_{pm}(g) = 2 (N Z)_{pm}, \ \ \ \  K^{(c)}_{pm}(g) = 4 (P N Z Q N Z)_{pm}  \\
&&  b(g) =- (N + 4 P N Z Q N) a^{\textrm{T}} - N \bar a^{\textrm{T}}
- 4 ( N Z P + 4 P N Z Q N Z P) t^{\textrm{T}}  \, , \nonumber
\end{eqnarray}
with
\begin{eqnarray}
&&
N= \textrm{diag} (1,2,3, \dots), \ \ \ P=  \textrm{diag} (0,1,0,1 \dots), \ \ \ Q=\textrm{diag} (1,0,1,0 \dots), \nonumber \\
&& t = (1,0,1,0, \dots), \ \ \ a=(a_1,0,a_3,0, \dots) , \ \ \ \bar
a=(0,\bar a_2,0,\bar a_4, \dots).
\end{eqnarray}
At least formally, the solution may be written as
\begin{eqnarray}
\label{sysnum} s(g) \ = \ ({\mathcal I}+K^{(m)}+2 K^{(c)})^{-1}\, b
\, ,
\end{eqnarray}
where ${\mathcal I}$ is the identity matrix. This similarity with \cite{BBKS} 
is crucial for what concerns the numerical treatment, in that everything 
works well (as in \cite{BBKS}), because (the matrix elements are the same and) 
the forcing elements have a similar behaviour for large $g$.

In the following we will be interested in a numerical analysis of (\ref{sysnum}), 
emphasising the deviation of the strong coupling behaviour of the first few $s_m(g)$ from 
the exact expansion of the previous Section. In this respect, a sufficient accuracy is achieved
upon truncating at $m=30$: this restriction still yields reliable results up to about $g \simeq 20$.

\subsection{The strong coupling corrections and the string side}

\begin{figure}[ht] \setlength{\unitlength}{1mm}
\begin{picture}(100,104)(-3,0)
\put(0,0){\includegraphics[width=0.95\linewidth]{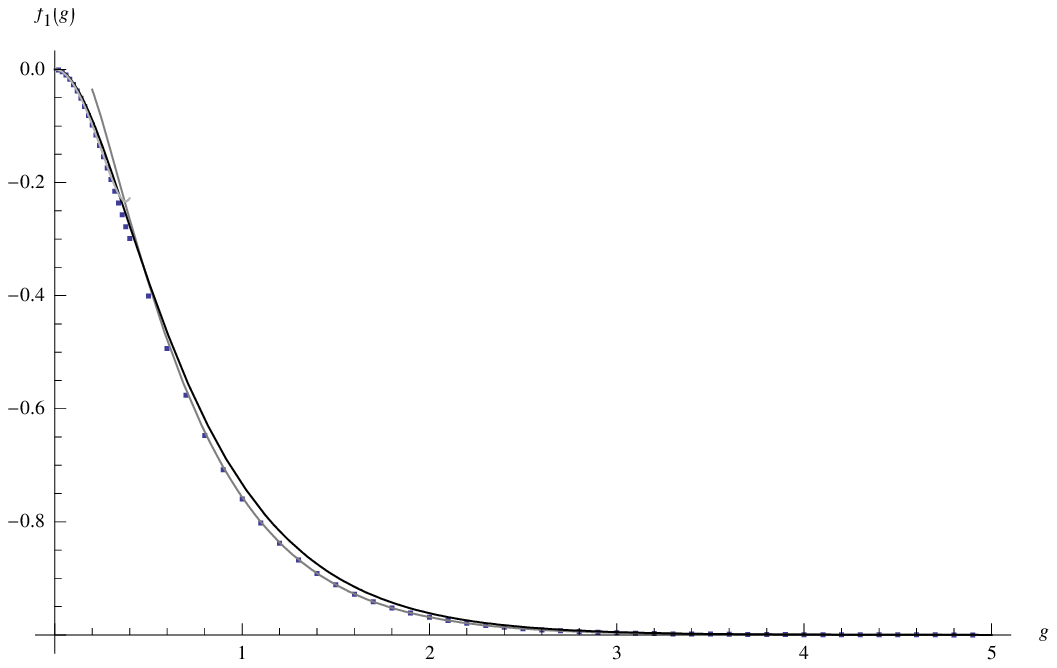}}
\put(139,40){\color{white}\rule{4.5mm}{40mm}}
\put(51,30){\includegraphics[width=85mm]{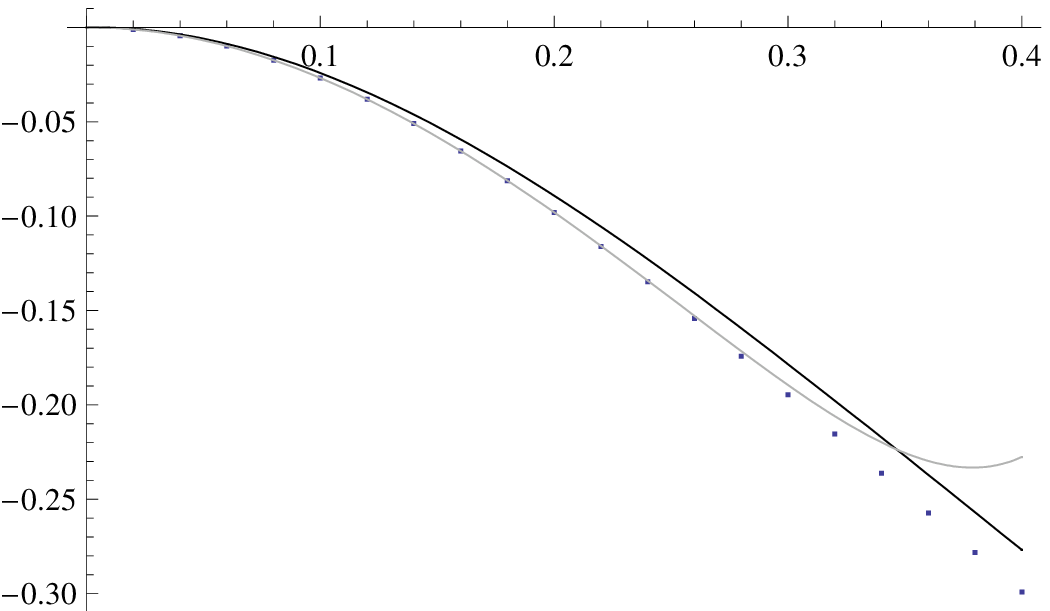}}
\put(11,60){\drawline(0,0)(10,0)(10,25)(0,25)(0,0)}
\put(21,85){\dottedline{3}(0,0)(30,-6)}
\put(21,60){\dottedline{3}(0,0)(30,-26)}
\end{picture}
\caption{\label{fig1} Comparison between numerical solution (dots),
fit as in formula (\ref{f1exp}) (dark grey line) and integral
representation (\ref{f1int}) (black line). Detail in the small
window: comparison between numerical solution (dots), weak coupling
expansion of \cite{FRS, BFR2}  (light grey line)  and integral
representation (\ref{f1int}) (black line).}
\end{figure}

Since the asymptotic expansion for any $s_{p}(g)$ truncates 
significantly at the (inverse) power $1/g^p$, the additional (non-perturbative) 
exponential corrections are simply given by 
subtracting this to the numerical solution of (\ref{sysnum}) \footnote{At the moment, we
observe the same exponential behaviour, with different constant factors (cf. below), 
for the first few $s_n(g)$, $n=1, \dots,5$.}. At any rate, our main concern here is 
$s_1(g)$ by virtue of its relation (\ref{s1f1}) to the generalised scaling function $f_1(g)$, 
whose asymptotic expansion $f(g)\doteq -1$ is in perfect agreement with the string energy density 
$\Delta-s=\gamma+L$ (cf. (\ref{Delta})), as perturbatively expanded
in \cite{FTT, AM}, because at present it does not seem to have the term $L$. 
Moreover, beyond this asymptotics there are additional, non-perturbative, 
exponential corrections, whose leading order (see below (\ref{f1exp})) 
is reported in figure \ref{fig1},
where the large $g$, asymptotic behaviour is already reached for $g \sim 3.5$.

We can then proceed to a more quantitative analysis. At first, 
we perform a fit of the numerical evaluation of
(\ref{sysnum}), in order to check whether it is compatible with the
proposed exact asymptotic expansion (\ref{sm}). This
analysis -- performed at the moment on the first few $s_n(g)$, $n=1, \dots,5$ --
rules out the possibility that other terms with the form $1/g^k$
might appear. Then, we find reasonable to conjecture a non-perturbative,
exponential term coming into play. In fact, a simple dimensional
consideration has led the authors of \cite{AM} to claim that,
regarding $f_1(g)$, the additional string energy density
\footnote{This has to be summed to $f(g) \ln s$ and is due to the
$SO(6)$ charge $j$.} is simply proportional to ($j$ and) the $O(6)$
sigma model mass gap, $m$, or in formul\ae \, $(f_1(g)+1) j=c m(g) j$
as long as $j\ll m$ (still $g\rightarrow +\infty$). \footnote{We may
expect that $m(g)$ in this formula differs from the mass gap as soon
as $j \ll g$ is no longer valid.} As a consequence, we use the
functional form implied by the mass gap formula (\cite{AM} and
reference [13] therein), i.e.
\begin{eqnarray}
f_1(g) = -1 + \kappa \, g^{1/4} \, e^{- \frac{\pi}{\sqrt{2}} \, g},
\ \ \ g \to \infty \, , \label {f1exp}
\end{eqnarray}
with only $\kappa$ as a free parameter to fit, and we find a perfect
agreement with the data in the region $g \in [3,20]$, resulting in
the best fit estimate $\kappa = 2.257 \pm 0.009$. On the other hand,
the $O(6)$ small coupling calculations of \cite{AM} entail the exact
value $\kappa= 2^{5/8} \pi^{1/4} c / \Gamma(5/4)=(2.265218666...)\, c$,
which compels us to the natural prediction that $c=1$. \footnote{In
this respect there is also an ongoing analytic evaluation.}

We can now comment a little about the integral summation
(\ref{f1int}) to $\tilde f_1(g)$. Despite the fact that it is able 
to qualitatively capture the main features of
the numerical solution for a wide range of $g$, it still fails to 
reproduce both the weak-coupling expansion, and the power factor $g^{\frac{1}{4}}$ 
of the non-perturbative term. We have a clear picture of this situation
in fig.~\ref{fig1}. Finally, we shall remark the perfect
agreement between the numerical solution and the first terms of the
weak-coupling expansion computed in \cite{FRS, BFR2}.

\section{Some final comments}

Unlike all the genuine (i.e. infinite) asymptotic series involved in the derivation 
of the leading term $f(g)$ (see, for instance, \cite{BKK}), which amounts to 
twice the cusp anomalous dimension \cite{KM}, all 
the analogous series for $f_1(g)$ truncate, thus giving rise to polynomials in $1/g$. 
Moreover, the latter become non-trivially corrected by exponentially small contributions. 
At this stage, a deeper comparison with string results (\cite{FTT} and references 
therein) is really compelling, but still demanding because of the summation of all the 
logarithmic contributions in the string loop expansion \cite{FTT}. Yet, this is the direct 
consequence of an ordering of different limits and the present results 
shed some light on what happens to the higher $f_n$, $n=2,3,\dots$ \, .


{\bf Acknowledgments} D.F. thanks D. Bombardelli, F. Buccheri, E.
Caliceti, F. Cannata, F. Ravanini, R. Soldati for
discussions, the INFN grant "Iniziativa specifica PI14" and the
international agreement INFN-MEC-2008 for travel financial support.
P.G. work is partially supported by MEC-FEDER (grant FPA2005-00188),
by the Spanish Consolider-Ingenio 2010 Programme CPAN
(CSD2007-00042) and by Xunta de Galicia (Conseller\'\i a de
Educaci\'on and grant PGIDIT06PXIB296182PR). M.R. thanks
INFN-Bologna for kind hospitality.

\end{document}